\documentclass[%
 aip,
 amsmath,amssymb,
 reprint,%
]{revtex4-1}

\usepackage{graphicx}
\usepackage{dcolumn}
\usepackage{bm}

\usepackage[utf8]{inputenc}
\usepackage[T1]{fontenc}
\usepackage{mathptmx}
\usepackage{changes}

\begin{document}

\preprint{AIP/123-QED}

\title{Probing the ultrafast gain and refractive index dynamics of a VECSEL}

\author{C. Kriso}
 \email{christian.kriso@physik.uni-marburg.de}
\author{T. Bergmeier}%
\affiliation{ 
Department of Physics and Material Sciences Center, Philipps-Universität Marburg, Renthof 5, 35032 Marburg, Germany
}%
\author{N. Giannini}
\author{A. R. Albrecht}
\author{M. Sheik-Bahae}
\affiliation{ 
Department of Physics and Astronomy, University of New Mexico, 210 Yale Blvd. NE, Albuquerque, New Mexico 87131, USA
}%
\author{S. Benis}
\author{S. Faryadras}
\author{E. W. Van Stryland}
\author{D. J. Hagan}
\affiliation{%
CREOL, The College of Optics and Photonics, University of Central Florida, Orlando, FL 32816, USA
}%
\author{M. Koch}
\author{G. Mette}
\author{A. Rahimi-Iman}
 \email{a.r-i@physik.uni-marburg.de}
\affiliation{ 
Department of Physics and Material Sciences Center, Philipps-Universität Marburg, Renthof 5, 35032 Marburg, Germany
}%
\date{\today}

\begin{abstract}
Typically, strong gain saturation and gain dynamics play a crucial role in semiconductor laser mode-locking. While there have been several investigations of the ultrafast gain dynamics in vertical-external-cavity surface-emitting lasers (VECSELs), little is known about the associated refractive index changes. Yet, such refractive index changes do not only have a profound impact on the pulse formation process leading to self-phase modulation, which needs to be compensated by dispersion, but they are also of particular relevance for assessing the feasibility of Kerr-lens mode-locking of VECSELs. Here, we measure both refractive index as well as gain dynamics of a VECSEL chip using the ultrafast beam deflection method. We find that, in contrast to the gain dynamics, the refractive index dynamics is dominated by an instantaneous ($\sim$100~fs) and a very slow component ($\sim$100~ps). The time-resolved measurement of nonlinear refraction allows us to predict a pulse-length dependent, effective nonlinear refractive index $n_{2,eff}$, which is shown to be negative and in the order of $10^{-16}$ $m^2/W$ for short pulse lengths ($\sim$100~fs)
. It becomes positive for large excitation fluences and large pulse lengths (few ps). These results agree with some previous reports of self-mode-locked VECSELs for which the cavity design and pulse properties determine sign and strength of the nonlinear refractive index when assuming Kerr-lens mode-locking.
\end{abstract}

\maketitle

Ultrashort-pulse mode-locking and frequency combs have emerged as a major topic in semiconductor laser research, and state-of-the-art semiconductor disk laser systems (or VECSELs) are under development that utilize passive mode-locking with various applications in mind such as spectroscopy, biomedicine, nonlinear optics and many more.\cite{Link2017, Voigt2017, Bek2015} Mostly optically-pumped,  VECSELs combine flexible wavelength design of the gain chip, high optical output powers and excellent beam quality.\cite{Kuznetsov1997,Rahimi-Iman2016, Guina2017} The external cavity allows to further functionalize the laser emission, for example to achieve single-frequency generation or nonlinear frequency conversion by inserting optical filters or nonlinear crystals into the cavity, respectively.\cite{Zhang2014, Scheller2010}

Mode-locking can routinely be achieved by inserting semiconductor-saturable absorber mirrors (SESAMs) into the cavity with the possibility of obtaining ultrashort pulses in the sub-100~fs regime as well as peak powers of several kilowatts.\cite{Waldburger2016, Laurain2018, Wilcox2013}

In this context, recently also saturable-absorber-free mode-locking, usually referred to as "self-mode-locking", has received considerable attention and has been demonstrated by several groups.\cite{Chen2011, Kornaszewski2012, Albrecht2013, Gaafar2014a, Bek2017} These results remain subject of ongoing debate within the community concerning what the driving mechanisms behind this phenomenon are, and whether those truly lead to a mode-locked state of the laser.\cite{Wilcox2013b, Gaafar2016, Escoto2020} Both, mode-locking by a four-wave-mixing nonlinearity in the gain chip\cite{Escoto2020} and Kerr-lens mode-locking have been discussed as possible explanations.\cite{Albrecht2013} In particular, the latter hypothesis triggered considerable efforts to characterize the nonlinear refractive index of the gain chip under realistic conditions, i.e. using probe irradiances and excitation fluences comparable to how they occur in a mode-locked VECSEL.\cite{Quarterman2015, Shaw2016, Quarterman2016, Kriso2019, Kriso2020} However, most of these investigations have been performed using pulse lengths of only a few hundreds of femtoseconds. Yet, pulses generated by self-mode-locked VECSELs usually are longer, that is in the few-ps regime down to sub-ps pulse durations. \cite{Gaafar2016} Therefore, considering  the strong gain dynamics of semiconductor lasers, which intrinsically affects the refractive index of the gain chip, such nonlinear lensing investigations so far have not provided a very accurate picture of the nonlinear refractive index of the gain chip. Time-resolved measurements of the refractive index dynamics of a VECSEL would therefore allow us to obtain a more realistic estimate of the strength of nonlinear lensing in VECSELs. Beyond this, it would allow us to obtain insight into the phase dynamics that affects a short pulse in VECSEL mode-locking. In phenomenological modeling of pulse formation, this is usually taken into account by introducing a constant, the linewidth enhancement factor, which relates gain changes to phase changes.\cite{Paschotta2002, Sieber2013} However, this is a strongly simplifying assumption and therefore the full knowledge of the refractive index dynamics might improve modeling significantly.

In this work, we measure the time-resolved nonlinear optical response of a gain chip using the recently developed ultrafast beam-deflection technique.\cite{Ferdinandus2013} This uniquely enables us to simultaneously acquire gain as well as refractive index dynamics. Moreover, we validate the results by additional Z-scan measurements well known from effective nonlinear lensing characterization of materials without temporal resolution.\cite{Sheik-Bahae1990} 

\begin{figure}
\includegraphics[width=8cm]{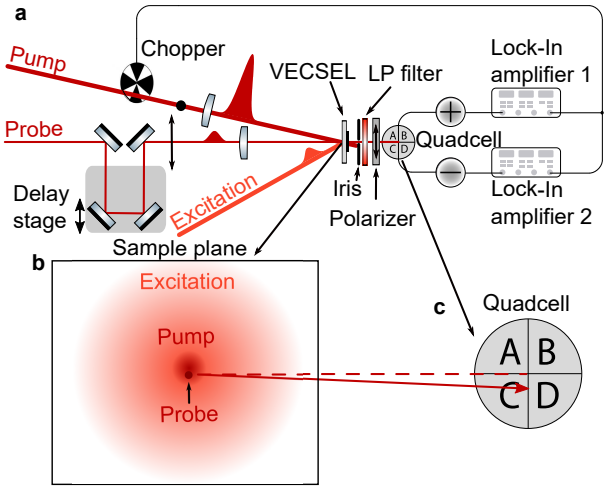}
\caption{\label{fig:setup} (a) Beam-deflection setup for measuring time-resolved transmission and refractive index changes of the VECSEL. A prepulse (at 780~nm), arriving 100~ps before the pump-probe measurement (at 1150~nm) takes place, excites the sample. Two lock-in amplifiers are used to measure simultaneously the sum \mbox{("+")} and difference \mbox{("-")} signal  of the segmented detector ("Quadcell"). (b) View on the sample (VECSEL) surface displaying the approximate size ratio and relative position of the spots of excitation as well as pump and probe beam. (c) The pump-induced refractive index change in the sample will lead to a deflection of the probe beam that will be detected by a non-zero difference signal between the upper and lower half of the segmented detector.}
\end{figure}

Our pump-probe setup, including a prepulse for excitation of the sample, is displayed in Fig.~\ref{fig:setup}(a). 
 Here, the probe spot is overlapped with the pump spot on the sample in a way that it sits exactly on the position of the largest slope of the Gaussian-shaped intensity profile of the pump as shown in Fig. \ref{fig:setup}(b) (which requires that the probe spot diameter is considerably smaller than the pump at the position of the sample). Consequently, the refractive index changes induced by the pump in the sample will lead to a deflection of the probe which is proportional to the refractive index change of the sample and can be detected by a segmented photodetector. In our scheme, we use lock-in detection to measure both pump-induced transmission changes by recording the sum-signal of the segmented detector and pump-induced deflection changes by recording the difference signal of the segmented detector. When varying the delay between pump and probe pulse, we can record time-resolved changes of transmission and deflection. For the beam deflection measurements, we used a 200~kHz femtosecond laser system (Light Conversion Carbide) pumping two optical parametric amplifiers (OPAs). With one OPA \mbox{(Orpheus-F)}, we generated  pulses at a center wavelength of 1163~nm and, with another OPA \mbox{(Orpheus-N-2H)}, pulses at a center wavelength of 780~nm. The latter is used to excite the sample with a prepulse arriving at the sample about 100~ps before the pump-probe measurement takes place. For the pump-probe measurements, a filter with center wavelength of 1150~nm and a bandwidth of 25~nm was used to align the laser spectrum with respect to the photoluminesence of the VECSEL sample (shown in the Supplementary Material). The pump and probe beams are obtained by splitting the 1150~nm laser beam with a 90:10-beam splitter. A half-wave plate and calcite polarizers are used to ensure good orthogonal polarization of pump and probe with respect to each other. The probe beam is attenuated by neutral-density filters in order to prevent detector saturation. To block the pump from reaching the detector, an iris and a polarizer are used which only transmit the probe. The half width (HW)1/$e^2$ of the probe beam at its focus and of the pump spot at the same position is 29~µm and 97~µm, respectively. Both, the pump-induced transmission and deflection signal scale approximately linearly when increasing the pump irradiance, as was verified with reference samples (see the Supplementary Material). The spot size of the excitation is made a lot larger (with a HW1/e$^2$ of about 500~µm on the chip) to ensure an approximately homogeneous sample excitation at the area where the pump-probe measurement takes place.  A long-pass filter is inserted after the sample to prevent the excitation beam at 780~nm from scattering into the detector. \\ 
The VECSEL sample used in this investigation consists of a resonant periodic gain structure of 10 InGaAs quantum wells separated by GaAsP barriers for strain compensation, and was grown by metalorganic chemical vapor deposition (MOCVD). Two InGaP layers of approximately 190 nm thickness surround the structure for charge carrier confinement. The sample used here is a distributed Bragg reflector (DBR)-free VECSEL or membrane external-cavity surface emitting laser (MECSEL). The GaAs growth substrate is etched away and the structure is van-der-Waals-bonded onto a 350 $\mu$m thick 4H-SiC heat spreader. MECSELs have been shown to exhibit an extraordinarily large wavelength tuning range and excellent thermal properties.\cite{Yang2015, Kahle2016, Yang2018} In this work, it enables us to conduct the beam deflection measurements in transmission geometry rather than in reflection which greatly simplifies the experiment. 

\begin{figure}
\includegraphics[width=8cm]{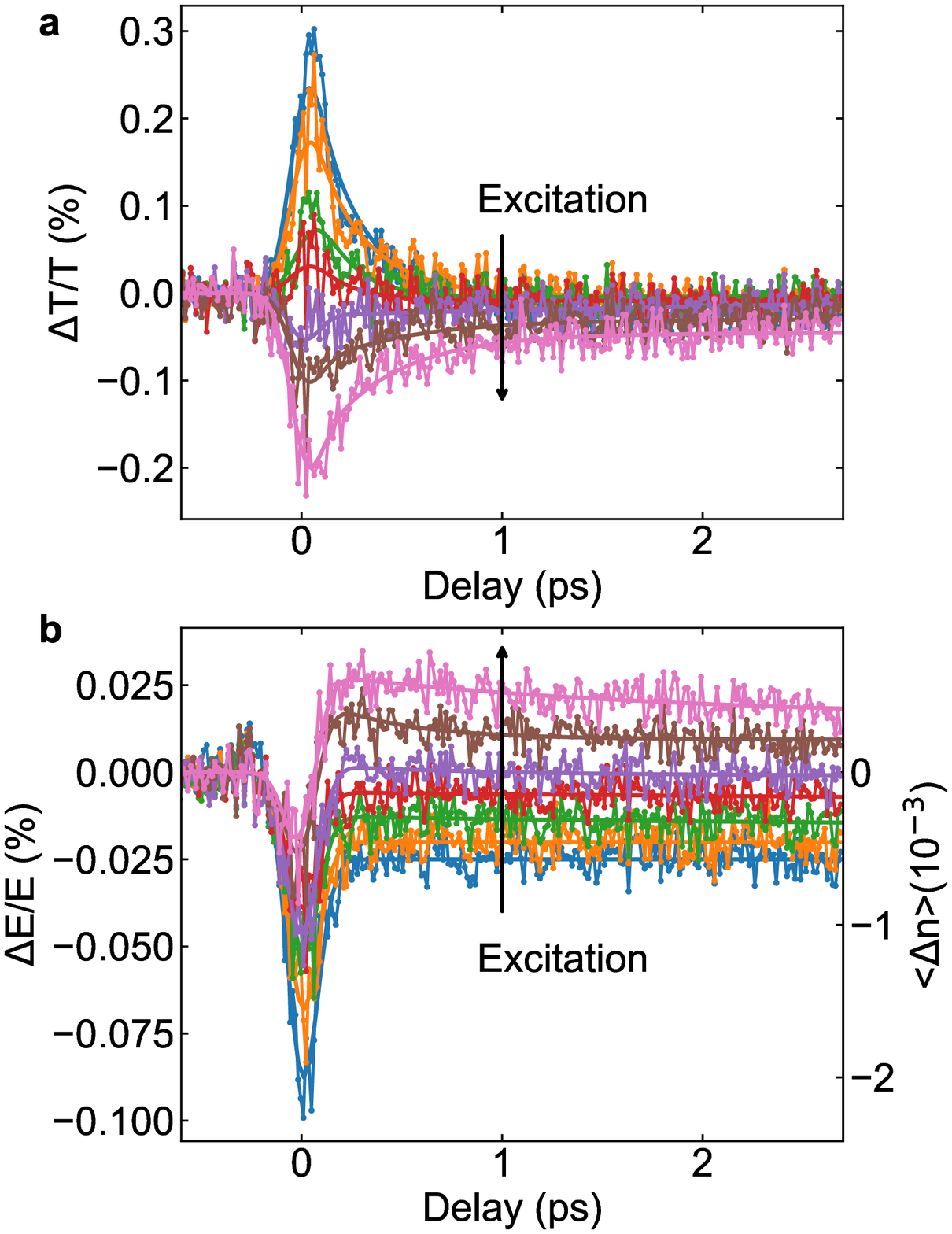}
\caption{\label{fig:Pump_probe} (a) Normalized probe transmission and (b) deflection of VECSEL for a pump irradiance of 1~GW/cm$^2$ and various excitation fluences (1.1~µJ/cm$^2$, 2.8~µJ/cm$^2$, 6.0~µJ/cm$^2$, 8.7~µJ/cm$^2$, 10.2~µJ/cm$^2$, 13.0~µJ/cm$^2$ and 21.2~µJ/cm$^2$ in the direction of the arrow). The probe beam experiences increased transmission when the pump experiences absorption and opposite behavior in the case of gain. The solid line displays the model used to fit the data (Eq. \ref{eq:pump_probe_eq}). The right axis in (b) displays the refractive index change averaged over the width of the cross-correlation of pump and probe pulse ($\sim$180~fs FWHM) that corresponds to the measured deflection.}  
\end{figure}
In the following, we investigate beam deflection measurements for different excitation fluences ranging from 1.1~µJ to 21.2~µJ and a fixed pump peak irradiance (1~GW/cm$^2$). Figure \ref{fig:Pump_probe}(a) and (b) show the normalised pump-induced transmission changes $\Delta T/T$ and the deflection changes $\Delta E/E$ as a function of the delay between pump and probe pulse, respectively.
Transmission changes $\Delta T$ are normalized with respect to the total signal of the probe in the absence of the pump, $T$, and deflection changes $\Delta E$ are normalized with respect to $E=T+\Delta T$, which also takes into account the transmission changes of the probe during the deflection measurement.

When the sample is only weakly excited, the probe pulse experiences first an increase of transmission before decaying back to a transmission change of close to zero. The deflection signal experiences a very fast negative response and subsequently a rather constant negative deflection signal over the few-ps time scale. Negative deflection corresponds to a negative refractive index change.

When the sample is strongly excited, the probe experiences first a strong decrease in transmission and subsequently a recovery to a slightly negative transmission change. This offset from zero transmission change at longer times indicates that the pump beam experiences net gain. Note that the probe beam will experience increased absorption when the pump experiences gain and opposite behavior in the case of absorption.\cite{Hall1990} The deflection signal becomes positive for long time scales and large excitation fluences. \\ 
Both, transmission and deflection measurements can be modeled by a response function of the form, 
\begin{equation}
R(t) = (b_0e^{-\frac{t}{\tau_0}}+b_1e^{-\frac{t}{\tau_1}})\theta(t)+b_2\delta(t),
\label{eq:response_function}
\end{equation}
where $\theta(t)$ and $\delta(t)$ are the Heaviside and Dirac delta function, respectively. The time constant $\tau_0$ models the non-instantaneous response of the sample consisting of carrier cooling in the absorption regime and carrier heating in the gain regime that occur with a time constant of several hundreds of fs. The time constant $\tau_1$ is usually in the order of magnitude of hundreds of ps or few ns and describes carrier relaxation or refilling.\cite{Mork1994}  The instantaneous response described by the Dirac delta function models both contributions from two-photon absorption or the ultrafast Kerr effect as well as relaxation of the excited carriers into a Fermi-Dirac distribution in the absorption regime or filling of the spectral holes in the gain regime, which is too fast to be resolved by the measurement conducted with a pump and probe pulse with a full-width at half maximum (FWHM) of 130~fs, respectively.

This response function is fitted to the experimental traces by 
\begin{equation}
\frac{\int_{-\infty}^{+\infty}\int_{-\infty}^{+\infty}R(t-t')I_{pu}(t') dt'I_{pr}(t-\tau)dt}{\int_{-\infty}^{+\infty}I_{pr}(t)dt}.
\label{eq:pump_probe_eq}
\end{equation}
Here, $I_{pu}$ and $I_{pr}$ represent the intensity envelope of the pump and probe pulse, respectively.

When fitting the pump-probe measurement of the transmission changes with Eqs.~\ref{eq:response_function} and \ref{eq:pump_probe_eq} one obtains a time constant $\tau_0$ of 300-400 fs for carrier cooling, i.e. when the sample is in the absorption regime and only weakly excited. This corresponds to values measured in Ref. \cite{Alfieri2017}. When the sample is strongly excited, this time constant increases slightly to 400-500 fs, corresponding to the time the carrier-distribution heats up to the lattice temperature. In comparison to the measurement of Ref. \cite{Baker2015}, the ultrafast gain recovery also contains a pronounced  instantaneous component ($\sim$100~fs) for large excitation fluences.

Interestingly, the effect of carrier cooling/heating, i.e. the component with a sub-ps time constant, is not very strong in the measurement of the deflection signal (Fig. \ref{fig:Pump_probe}(b)), as that signal mostly consists of an ultrafast ($\sim$100~fs) and a very slow component ($\sim$100~ps). In contrast to measurements of semiconductor optical amplifiers, the instantaneous negative decrease of the refractive index reduces significantly with increased excitation fluence, while otherwise the trend is similar.\cite{Hall1990,Mork1995} 

The time-resolved deflection measurement can be mapped to the refractive index change averaged over the length of the probe pulse, $<\Delta n( \tau )>$, where $\tau$ represents the delay between pump and probe pulse. This is done by comparing the deflection signal of the VECSEL to the deflection signal of a reference sample (SiC) with a known nonlinear refractive index $n_2$ (see the Supplementary Material for more details of this procedure). The right axis of Fig. \ref{fig:Pump_probe}(b) displays the corresponding $<\Delta n( \tau )>$ for the measurement of the VECSEL sample. 

\begin{figure}[h!]
\includegraphics[width=6.5cm]{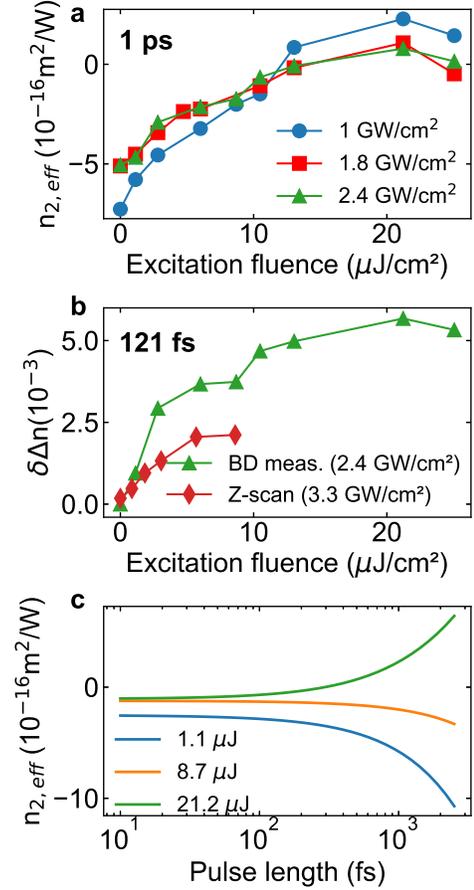}
\caption{\label{fig:pulse_length_dependent_n2_eff} (a) Effective, excitation-dependent, nonlinear refractive index n$_{2,eff}$ calculated for a pulse length (FWHM) of 1~ps and several pump peak irradiances. (b) Excitation-induced refractive index change $\delta \Delta n$ obtained from a Z-scan measurement conducted with a peak irradiance of 3.3~GW/cm$^2$ and a pulse length (FWHM) of 121~fs and the beam deflection measurement performed with 2.4~GW/cm$^2$ pump peak irradiance.  (c) Calculated n$_{2,eff}$ as a function of the pulse length (FWHM) for different excitation fluences and a pump peak irradiance of 1~GW/cm$^2$. The calculation of $n_{2,eff}$ (Eq.~\ref{eq:n_2_eff}) is based on the fitted response function of the nonlinear refractive index change.}
\end{figure}

The response function, obtained by fitting $<n(\tau)>$ to Eq.~\ref{eq:pump_probe_eq}, can be used to calculate the effective nonlinear refractive index $n_{2,eff}$. This quantity describes the nonlinear refractive index that would be seen by a single beam propagating through the sample, for example when performing a Z-scan measurement, and depends on the pulse length. It relates to the total refractive index change by $\Delta n=n_{2,eff}I$, with $I$ being the irradiance of the single beam. It can be calculated by\cite{Reichert2014}
\begin{align}
n_{2,eff} = \frac{\int_{-\infty}^{+\infty}\int_{-\infty}^{+\infty}R(t-t')I(t') dt'I(t)dt}{\int_{-\infty}^{+\infty}I^2(t)dt}
\label{eq:n_2_eff}.
\end{align}
Here, $I(t)$ is the intensity pulse  envelope of the single beam.
Fig. \ref{fig:pulse_length_dependent_n2_eff}(a) shows the excitation-dependent $n_{2,eff}$ calculated from various deflection measurements with different pump irradiances for a pulse length of 1~ps. One can see that $n_{2,eff}$ is similar for all pump irradiances, being around \mbox{$-$6$\cdot$10$^{-16}~$m$^2$/W} for no excitation and increasing to slightly positive values of around \mbox{$+$2$\cdot$10$^{-16}~$m$^2$/W} for large excitation fluences. The fact that $n_{2,eff}$, calculated for 1~ps, is approximately independent of the pump irradiance demonstrates the third-order character of the nonlinearity for both the instantaneous ($\sim$100~fs) and the slower component ($\sim$1~ps). 

In addition to the beam deflection measurements, Z-scan measurements were performed on the unexcited sample, which yield an $n_{2,eff}$ of $-$1.2$\cdot$10$^{-16}$~m$^2$/W. This corresponds in sign and order of magnitude to the $n_{2,eff}$ of around $-$3$\cdot$10$^{-16}$~m$^2$/W calculated from the beam deflection measurements with the Z-scan pulse length of 121~fs. Furthermore, the trend of the refractive index change with increasing excitation was measured by Z-scan and is compared in Fig.~\ref{fig:pulse_length_dependent_n2_eff}(b) to the beam deflection measurement with relatively good agreement of both measurements. For details of the measurement of the excitation-induced refractive index change $\delta \Delta n$, we refer to the Supplementary Material. Also, we note that  both the order of magnitude of $n_{2,eff}$ and the trend with increasing excitation correspond to previous measurements of nonlinear refraction in VECSELs as well as theoretical investigations.\cite{Shaw2016, Quarterman2016, Kriso2019, Kriso2020, Sheik-Bahae1994} 

We proceed to calculate the pulse-length dependent n$_{2,eff}$ for several excitation fluences as shown in Fig. \ref{fig:pulse_length_dependent_n2_eff}(c). It can be seen that, for low excitation fluences, $n_{2,eff}$ is negative for all pulse lengths and increases strongly in magnitude for pulse lengths larger than 1~ps. This is a consequence of the nearly constant refractive index change at longer delays as shown in Fig. \ref{fig:Pump_probe}(b). In contrast, for large excitation fluences, $n_{2,eff}$ changes sign from negative to positive when going to pulse lengths beyond 1000~fs. 

It is interesting to compare the trend of the pulse-width-dependent, as well as the excitation-dependent, effective nonlinear refractive index with previous observations of self-mode-locking in VECSELs. In Ref.\cite{Albrecht2013}, self-mode-locking in a V-cavity was shown to possibly originate from a negative nonlinear refractive index in the order of 10$^{-16}$~m$^2$/W. Interestingly, when increasing the pump power, the measured pulse width decreased from >1~ps  to <500~fs which probably guarantees a negative effective nonlinear refractive index over the whole excitation range as shown by our investigations. 

In Ref. \cite{Rahimi-Iman2016b}, self-mode-locking was reported for a linear cavity. The insertion of a slit in front of the outcoupling mirror allows only assuming Kerr-lens mode-locking with a nonlinear lens of positive focal length. As the pulse length obtained in this experiment was 3.5~ps, a positive Kerr lens is indeed expected and thus the assumption of Kerr-lens mode-locking caused by a nonlinear refractive index in the order of 10$^{-16}$~cm$^2$/W is justified. 

Beyond assessing the feasibility of Kerr-lens mode-locking of VECSELs, transient nonlinear refractive index changes generally play a crucial role in modeling pulse formation of VECSELs as it causes chirped pulses. In the semi-classical approach used in Refs. \cite{Paschotta2002, Sieber2013} to model SESAM-mode-locking of VECSELs, the refractive index change induced by the changes in carrier occupation is modeled by the so-called linewidth-enhancement factor $\alpha$, that relates transient pulse phase changes $\Delta \varphi(t)$ to the time-dependent gain $g(t)$,\cite{Paschotta2002}
\begin{equation}
\Delta \varphi(t)=-\alpha g(t)/2.
\label{eq:lef}
\end{equation}
However, our measurements show that the refractive index dynamics, which directly translates to the pulse phase changes via $\Delta \varphi(t)=k\Delta n(t)$, with $k$ being the wave vector, differs significantly in shape and relative contributions of components with different time scales from the gain dynamics. Therefore, it appears useful to incorporate the response function of the refractive index measured here into pulse simulations for more realistic modeling instead of using the phenomenological parameter $\alpha$. 

In conclusion, we have probed both the gain and refractive index dynamics of a VECSEL chip under conditions similar to laser operation. Our results allow us to retrieve the response function of the refractive index change and to predict a pulse-length dependent effective nonlinear refractive index, which is negative and in the order of 10$^{-16}$~m$^2$/W for sub-ps pulse lengths but becomes positive for large excitation fluences and ps pulse lengths. These findings support the assumption of Kerr-lens mode-locking for some self-mode-locking results obtained with VECSELs. Additionally, our results might improve modeling of pulse formation in VECSELs by providing the time-resolved refractive index response of a VECSEL in the gain regime that could be directly incorporated into pulse-shaping simulations instead of a constant linewidth-enhancement factor. \\ \\
See the Supplementary Material for further information about the VECSEL characterization and additional beam deflection as well as Z-scan measurements\\ \\
The authors thank U. Höfer for providing the 200~kHz laser system. C.K. thanks all members of the Nonlinear Optics group at CREOL for outstanding support and useful discussions during his visit and acknowledges financial support by the Deutscher Akademischer Austauschdienst (DAAD). G.M. and T.B. gratefully acknowledge funding by the Deutsche Forschungsgemeinschaft (DFG) via Project-ID 223848855-SFB 1083 and GRK 1782. This work was funded by the DFG through project grants DFG RA 2841/1-1 and DFG RA 2841/1-3. This work was performed, in part, at the Center for Integrated Nanotechnologies, an Office of Science User Facility operated for the U.S. Department of Energy (DOE) Office of Science by Los Alamos National Laboratory (Contract 89233218CNA000001) and Sandia National Laboratories
(Contract DE-NA-0003525).

\section*{Data availability}
The data that support the findings of this study are available from the corresponding author upon reasonable request.
\bibliography{BD_VECSEL}

\begin{thebibliography}{39}%
\makeatletter
\providecommand \@ifxundefined [1]{%
 \@ifx{#1\undefined}
}%
\providecommand \@ifnum [1]{%
 \ifnum #1\expandafter \@firstoftwo
 \else \expandafter \@secondoftwo
 \fi
}%
\providecommand \@ifx [1]{%
 \ifx #1\expandafter \@firstoftwo
 \else \expandafter \@secondoftwo
 \fi
}%
\providecommand \natexlab [1]{#1}%
\providecommand \enquote  [1]{``#1''}%
\providecommand \bibnamefont  [1]{#1}%
\providecommand \bibfnamefont [1]{#1}%
\providecommand \citenamefont [1]{#1}%
\providecommand \href@noop [0]{\@secondoftwo}%
\providecommand \href [0]{\begingroup \@sanitize@url \@href}%
\providecommand \@href[1]{\@@startlink{#1}\@@href}%
\providecommand \@@href[1]{\endgroup#1\@@endlink}%
\providecommand \@sanitize@url [0]{\catcode `\\12\catcode `\$12\catcode
  `\&12\catcode `\#12\catcode `\^12\catcode `\_12\catcode `\%12\relax}%
\providecommand \@@startlink[1]{}%
\providecommand \@@endlink[0]{}%
\providecommand \url  [0]{\begingroup\@sanitize@url \@url }%
\providecommand \@url [1]{\endgroup\@href {#1}{\urlprefix }}%
\providecommand \urlprefix  [0]{URL }%
\providecommand \Eprint [0]{\href }%
\providecommand \doibase [0]{http://dx.doi.org/}%
\providecommand \selectlanguage [0]{\@gobble}%
\providecommand \bibinfo  [0]{\@secondoftwo}%
\providecommand \bibfield  [0]{\@secondoftwo}%
\providecommand \translation [1]{[#1]}%
\providecommand \BibitemOpen [0]{}%
\providecommand \bibitemStop [0]{}%
\providecommand \bibitemNoStop [0]{.\EOS\space}%
\providecommand \EOS [0]{\spacefactor3000\relax}%
\providecommand \BibitemShut  [1]{\csname bibitem#1\endcsname}%
\let\auto@bib@innerbib\@empty
\bibitem [{\citenamefont {Link}\ \emph {et~al.}(2017)\citenamefont {Link},
  \citenamefont {Maas}, \citenamefont {Waldburger},\ and\ \citenamefont
  {Keller}}]{Link2017}%
  \BibitemOpen
  \bibfield  {author} {\bibinfo {author} {\bibfnamefont {S.~M.}\ \bibnamefont
  {Link}}, \bibinfo {author} {\bibfnamefont {D.~J. H.~C.}\ \bibnamefont
  {Maas}}, \bibinfo {author} {\bibfnamefont {D.}~\bibnamefont {Waldburger}}, \
  and\ \bibinfo {author} {\bibfnamefont {U.}~\bibnamefont {Keller}},\
  }\bibfield  {title} {\enquote {\bibinfo {title} {{Dual-comb spectroscopy of
  water vapor with a free-running semiconductor disk laser}},}\ }\href@noop {}
  {\bibfield  {journal} {\bibinfo  {journal} {Science}\ }\textbf {\bibinfo
  {volume} {356}},\ \bibinfo {pages} {1164--1168} (\bibinfo {year}
  {2017})}\BibitemShut {NoStop}%
\bibitem [{\citenamefont {Voigt}\ \emph {et~al.}(2017)\citenamefont {Voigt},
  \citenamefont {Emaury}, \citenamefont {Bethge}, \citenamefont {Waldburger},
  \citenamefont {Link}, \citenamefont {Carta}, \citenamefont {van~der Bourg},
  \citenamefont {Helmchen},\ and\ \citenamefont {Keller}}]{Voigt2017}%
  \BibitemOpen
  \bibfield  {author} {\bibinfo {author} {\bibfnamefont {F.~F.}\ \bibnamefont
  {Voigt}}, \bibinfo {author} {\bibfnamefont {F.}~\bibnamefont {Emaury}},
  \bibinfo {author} {\bibfnamefont {P.}~\bibnamefont {Bethge}}, \bibinfo
  {author} {\bibfnamefont {D.}~\bibnamefont {Waldburger}}, \bibinfo {author}
  {\bibfnamefont {S.~M.}\ \bibnamefont {Link}}, \bibinfo {author}
  {\bibfnamefont {S.}~\bibnamefont {Carta}}, \bibinfo {author} {\bibfnamefont
  {A.}~\bibnamefont {van~der Bourg}}, \bibinfo {author} {\bibfnamefont
  {F.}~\bibnamefont {Helmchen}}, \ and\ \bibinfo {author} {\bibfnamefont
  {U.}~\bibnamefont {Keller}},\ }\bibfield  {title} {\enquote {\bibinfo {title}
  {{Multiphoton in vivo imaging with a femtosecond semiconductor disk
  laser}},}\ }\href@noop {} {\bibfield  {journal} {\bibinfo  {journal}
  {Biomedical Optics Express}\ }\textbf {\bibinfo {volume} {8}},\ \bibinfo
  {pages} {3213--3231} (\bibinfo {year} {2017})}\BibitemShut {NoStop}%
\bibitem [{\citenamefont {Bek}\ \emph {et~al.}(2015)\citenamefont {Bek},
  \citenamefont {Baumg{\"{a}}rtner}, \citenamefont {Sauter}, \citenamefont
  {Kahle}, \citenamefont {Schwarzb{\"{a}}ck}, \citenamefont {Jetter},\ and\
  \citenamefont {Michler}}]{Bek2015}%
  \BibitemOpen
  \bibfield  {author} {\bibinfo {author} {\bibfnamefont {R.}~\bibnamefont
  {Bek}}, \bibinfo {author} {\bibfnamefont {S.}~\bibnamefont
  {Baumg{\"{a}}rtner}}, \bibinfo {author} {\bibfnamefont {F.}~\bibnamefont
  {Sauter}}, \bibinfo {author} {\bibfnamefont {H.}~\bibnamefont {Kahle}},
  \bibinfo {author} {\bibfnamefont {T.}~\bibnamefont {Schwarzb{\"{a}}ck}},
  \bibinfo {author} {\bibfnamefont {M.}~\bibnamefont {Jetter}}, \ and\ \bibinfo
  {author} {\bibfnamefont {P.}~\bibnamefont {Michler}},\ }\bibfield  {title}
  {\enquote {\bibinfo {title} {{Intra-cavity frequency-doubled mode-locked
  semiconductor disk laser at 325 nm}},}\ }\href@noop {} {\bibfield  {journal}
  {\bibinfo  {journal} {Optics Express}\ }\textbf {\bibinfo {volume} {23}},\
  \bibinfo {pages} {19947--19953} (\bibinfo {year} {2015})}\BibitemShut
  {NoStop}%
\bibitem [{\citenamefont {Kuznetsov}\ \emph {et~al.}(1997)\citenamefont
  {Kuznetsov}, \citenamefont {Hakimi}, \citenamefont {Sprague},\ and\
  \citenamefont {Mooradian}}]{Kuznetsov1997}%
  \BibitemOpen
  \bibfield  {author} {\bibinfo {author} {\bibfnamefont {M.}~\bibnamefont
  {Kuznetsov}}, \bibinfo {author} {\bibfnamefont {F.}~\bibnamefont {Hakimi}},
  \bibinfo {author} {\bibfnamefont {R.}~\bibnamefont {Sprague}}, \ and\
  \bibinfo {author} {\bibfnamefont {A.}~\bibnamefont {Mooradian}},\ }\bibfield
  {title} {\enquote {\bibinfo {title} {{High-power ({\textgreater}0.5-W CW)
  diode-pumped vertical-external-cavity surface-emitting semiconductor lasers
  with circular TEM$_{00}$ beams}},}\ }\href@noop {} {\bibfield  {journal}
  {\bibinfo  {journal} {IEEE Photonics Technology Letters}\ }\textbf {\bibinfo
  {volume} {9}},\ \bibinfo {pages} {1063--1065} (\bibinfo {year}
  {1997})}\BibitemShut {NoStop}%
\bibitem [{\citenamefont {Rahimi-Iman}(2016)}]{Rahimi-Iman2016}%
  \BibitemOpen
  \bibfield  {author} {\bibinfo {author} {\bibfnamefont {A.}~\bibnamefont
  {Rahimi-Iman}},\ }\bibfield  {title} {\enquote {\bibinfo {title} {{Recent
  advances in VECSELs}},}\ }\href@noop {} {\bibfield  {journal} {\bibinfo
  {journal} {Journal of Optics}\ }\textbf {\bibinfo {volume} {18}},\ \bibinfo
  {pages} {093003} (\bibinfo {year} {2016})}\BibitemShut {NoStop}%
\bibitem [{\citenamefont {Guina}, \citenamefont {Rantam{\"{a}}ki},\ and\
  \citenamefont {H{\"{a}}rk{\"{o}}nen}(2017)}]{Guina2017}%
  \BibitemOpen
  \bibfield  {author} {\bibinfo {author} {\bibfnamefont {M.}~\bibnamefont
  {Guina}}, \bibinfo {author} {\bibfnamefont {A.}~\bibnamefont
  {Rantam{\"{a}}ki}}, \ and\ \bibinfo {author} {\bibfnamefont {A.}~\bibnamefont
  {H{\"{a}}rk{\"{o}}nen}},\ }\bibfield  {title} {\enquote {\bibinfo {title}
  {{Optically pumped VECSELs : review of technology and progress}},}\
  }\href@noop {} {\bibfield  {journal} {\bibinfo  {journal} {Journal of Physics
  D: Applied Physics}\ }\textbf {\bibinfo {volume} {50}},\ \bibinfo {pages}
  {383001} (\bibinfo {year} {2017})}\BibitemShut {NoStop}%
\bibitem [{\citenamefont {Zhang}\ \emph {et~al.}(2014)\citenamefont {Zhang},
  \citenamefont {Heinen}, \citenamefont {Wichmann}, \citenamefont
  {M{\"{o}}ller}, \citenamefont {Kunert}, \citenamefont {Rahimi-iman},
  \citenamefont {Stolz},\ and\ \citenamefont {Koch}}]{Zhang2014}%
  \BibitemOpen
  \bibfield  {author} {\bibinfo {author} {\bibfnamefont {F.}~\bibnamefont
  {Zhang}}, \bibinfo {author} {\bibfnamefont {B.}~\bibnamefont {Heinen}},
  \bibinfo {author} {\bibfnamefont {M.}~\bibnamefont {Wichmann}}, \bibinfo
  {author} {\bibfnamefont {C.}~\bibnamefont {M{\"{o}}ller}}, \bibinfo {author}
  {\bibfnamefont {B.}~\bibnamefont {Kunert}}, \bibinfo {author} {\bibfnamefont
  {A.}~\bibnamefont {Rahimi-iman}}, \bibinfo {author} {\bibfnamefont
  {W.}~\bibnamefont {Stolz}}, \ and\ \bibinfo {author} {\bibfnamefont
  {M.}~\bibnamefont {Koch}},\ }\bibfield  {title} {\enquote {\bibinfo {title}
  {{A 23-watt single-frequency vertical-external-cavity surface-emitting
  laser}},}\ }\href@noop {} {\bibfield  {journal} {\bibinfo  {journal} {Optics
  Express}\ }\textbf {\bibinfo {volume} {22}},\ \bibinfo {pages} {12817--12822}
  (\bibinfo {year} {2014})}\BibitemShut {NoStop}%
\bibitem [{\citenamefont {Scheller}\ \emph {et~al.}(2010)\citenamefont
  {Scheller}, \citenamefont {Yarborough}, \citenamefont {Moloney},
  \citenamefont {Fallahi}, \citenamefont {Koch},\ and\ \citenamefont
  {Koch}}]{Scheller2010}%
  \BibitemOpen
  \bibfield  {author} {\bibinfo {author} {\bibfnamefont {M.}~\bibnamefont
  {Scheller}}, \bibinfo {author} {\bibfnamefont {J.~M.}\ \bibnamefont
  {Yarborough}}, \bibinfo {author} {\bibfnamefont {J.~V.}\ \bibnamefont
  {Moloney}}, \bibinfo {author} {\bibfnamefont {M.}~\bibnamefont {Fallahi}},
  \bibinfo {author} {\bibfnamefont {M.}~\bibnamefont {Koch}}, \ and\ \bibinfo
  {author} {\bibfnamefont {S.~W.}\ \bibnamefont {Koch}},\ }\bibfield  {title}
  {\enquote {\bibinfo {title} {{Room temperature continuous wave milliwatt
  terahertz source}},}\ }\href@noop {} {\bibfield  {journal} {\bibinfo
  {journal} {Optics Expres}\ }\textbf {\bibinfo {volume} {18}},\ \bibinfo
  {pages} {27112--27117} (\bibinfo {year} {2010})}\BibitemShut {NoStop}%
\bibitem [{\citenamefont {Waldburger}\ \emph {et~al.}(2016)\citenamefont
  {Waldburger}, \citenamefont {Link}, \citenamefont {Mangold}, \citenamefont
  {Alfieri}, \citenamefont {Gini}, \citenamefont {Golling}, \citenamefont
  {Tilma},\ and\ \citenamefont {Keller}}]{Waldburger2016}%
  \BibitemOpen
  \bibfield  {author} {\bibinfo {author} {\bibfnamefont {D.}~\bibnamefont
  {Waldburger}}, \bibinfo {author} {\bibfnamefont {S.~M.}\ \bibnamefont
  {Link}}, \bibinfo {author} {\bibfnamefont {M.}~\bibnamefont {Mangold}},
  \bibinfo {author} {\bibfnamefont {C.~G.~E.}\ \bibnamefont {Alfieri}},
  \bibinfo {author} {\bibfnamefont {E.}~\bibnamefont {Gini}}, \bibinfo {author}
  {\bibfnamefont {M.}~\bibnamefont {Golling}}, \bibinfo {author} {\bibfnamefont
  {B.~W.}\ \bibnamefont {Tilma}}, \ and\ \bibinfo {author} {\bibfnamefont
  {U.}~\bibnamefont {Keller}},\ }\bibfield  {title} {\enquote {\bibinfo {title}
  {{High-power 100 fs semiconductor disk lasers}},}\ }\href@noop {} {\bibfield
  {journal} {\bibinfo  {journal} {Optica}\ }\textbf {\bibinfo {volume} {3}},\
  \bibinfo {pages} {844--852} (\bibinfo {year} {2016})}\BibitemShut {NoStop}%
\bibitem [{\citenamefont {Laurain}\ \emph {et~al.}(2018)\citenamefont
  {Laurain}, \citenamefont {Kilen}, \citenamefont {Hader}, \citenamefont
  {Perez}, \citenamefont {Ludewig}, \citenamefont {Stolz}, \citenamefont
  {Balakrishnan}, \citenamefont {Koch},\ and\ \citenamefont
  {Moloney}}]{Laurain2018}%
  \BibitemOpen
  \bibfield  {author} {\bibinfo {author} {\bibfnamefont {A.}~\bibnamefont
  {Laurain}}, \bibinfo {author} {\bibfnamefont {I.}~\bibnamefont {Kilen}},
  \bibinfo {author} {\bibfnamefont {J.}~\bibnamefont {Hader}}, \bibinfo
  {author} {\bibfnamefont {A.~R.}\ \bibnamefont {Perez}}, \bibinfo {author}
  {\bibfnamefont {P.}~\bibnamefont {Ludewig}}, \bibinfo {author} {\bibfnamefont
  {W.}~\bibnamefont {Stolz}}, \bibinfo {author} {\bibfnamefont
  {G.}~\bibnamefont {Balakrishnan}}, \bibinfo {author} {\bibfnamefont {S.~W.}\
  \bibnamefont {Koch}}, \ and\ \bibinfo {author} {\bibfnamefont {J.~V.}\
  \bibnamefont {Moloney}},\ }\bibfield  {title} {\enquote {\bibinfo {title}
  {{Modeling and experimental realization of modelocked VECSEL producing high
  power sub-100 fs pulses}},}\ }\href@noop {} {\bibfield  {journal} {\bibinfo
  {journal} {Applied Physics Letters}\ }\textbf {\bibinfo {volume} {113}},\
  \bibinfo {pages} {121113} (\bibinfo {year} {2018})}\BibitemShut {NoStop}%
\bibitem [{\citenamefont {Wilcox}\ \emph {et~al.}(2013)\citenamefont {Wilcox},
  \citenamefont {Tropper}, \citenamefont {Beere}, \citenamefont {Ritchie},
  \citenamefont {Heinen},\ and\ \citenamefont {Stolz}}]{Wilcox2013}%
  \BibitemOpen
  \bibfield  {author} {\bibinfo {author} {\bibfnamefont {K.~G.}\ \bibnamefont
  {Wilcox}}, \bibinfo {author} {\bibfnamefont {A.~C.}\ \bibnamefont {Tropper}},
  \bibinfo {author} {\bibfnamefont {H.~E.}\ \bibnamefont {Beere}}, \bibinfo
  {author} {\bibfnamefont {D.~A.}\ \bibnamefont {Ritchie}}, \bibinfo {author}
  {\bibfnamefont {B.}~\bibnamefont {Heinen}}, \ and\ \bibinfo {author}
  {\bibfnamefont {W.}~\bibnamefont {Stolz}},\ }\bibfield  {title} {\enquote
  {\bibinfo {title} {{4.35 kw peak power femtosecond pulse mode-locked VECSEL
  for supercontinuum generation}},}\ }\href@noop {} {\bibfield  {journal}
  {\bibinfo  {journal} {Optics Express}\ }\textbf {\bibinfo {volume} {21}},\
  \bibinfo {pages} {1599--1605} (\bibinfo {year} {2013})}\BibitemShut {NoStop}%
\bibitem [{\citenamefont {Chen}\ \emph {et~al.}(2011)\citenamefont {Chen},
  \citenamefont {Lee}, \citenamefont {Liang}, \citenamefont {Lin},
  \citenamefont {Su},\ and\ \citenamefont {Huang}}]{Chen2011}%
  \BibitemOpen
  \bibfield  {author} {\bibinfo {author} {\bibfnamefont {Y.~F.}\ \bibnamefont
  {Chen}}, \bibinfo {author} {\bibfnamefont {Y.~C.}\ \bibnamefont {Lee}},
  \bibinfo {author} {\bibfnamefont {H.~C.}\ \bibnamefont {Liang}}, \bibinfo
  {author} {\bibfnamefont {K.~Y.}\ \bibnamefont {Lin}}, \bibinfo {author}
  {\bibfnamefont {K.~W.}\ \bibnamefont {Su}}, \ and\ \bibinfo {author}
  {\bibfnamefont {K.~F.}\ \bibnamefont {Huang}},\ }\bibfield  {title} {\enquote
  {\bibinfo {title} {{Femtosecond high-power spontaneous mode-locked operation
  in vertical-external cavity surface-emitting laser with gigahertz
  oscillation}},}\ }\href@noop {} {\bibfield  {journal} {\bibinfo  {journal}
  {Optics Letters}\ }\textbf {\bibinfo {volume} {36}},\ \bibinfo {pages}
  {4581--4583} (\bibinfo {year} {2011})}\BibitemShut {NoStop}%
\bibitem [{\citenamefont {Kornaszewski}\ \emph {et~al.}(2012)\citenamefont
  {Kornaszewski}, \citenamefont {Maker}, \citenamefont {Malcolm}, \citenamefont
  {Butkus}, \citenamefont {Rafailov},\ and\ \citenamefont
  {Hamilton}}]{Kornaszewski2012}%
  \BibitemOpen
  \bibfield  {author} {\bibinfo {author} {\bibfnamefont {L.}~\bibnamefont
  {Kornaszewski}}, \bibinfo {author} {\bibfnamefont {G.}~\bibnamefont {Maker}},
  \bibinfo {author} {\bibfnamefont {G.~P.}\ \bibnamefont {Malcolm}}, \bibinfo
  {author} {\bibfnamefont {M.}~\bibnamefont {Butkus}}, \bibinfo {author}
  {\bibfnamefont {E.}~\bibnamefont {Rafailov}}, \ and\ \bibinfo {author}
  {\bibfnamefont {C.~J.}\ \bibnamefont {Hamilton}},\ }\bibfield  {title}
  {\enquote {\bibinfo {title} {{SESAM-free mode-locked semiconductor disk
  laser}},}\ }\href@noop {} {\bibfield  {journal} {\bibinfo  {journal} {Laser
  and Photonics Review}\ }\textbf {\bibinfo {volume} {6}},\ \bibinfo {pages}
  {20--23} (\bibinfo {year} {2012})}\BibitemShut {NoStop}%
\bibitem [{\citenamefont {Albrecht}\ \emph {et~al.}(2013)\citenamefont
  {Albrecht}, \citenamefont {Wang}, \citenamefont {Ghasemkhani}, \citenamefont
  {Seletskiy}, \citenamefont {Cederberg},\ and\ \citenamefont
  {Sheik-Bahae}}]{Albrecht2013}%
  \BibitemOpen
  \bibfield  {author} {\bibinfo {author} {\bibfnamefont {A.~R.}\ \bibnamefont
  {Albrecht}}, \bibinfo {author} {\bibfnamefont {Y.}~\bibnamefont {Wang}},
  \bibinfo {author} {\bibfnamefont {M.}~\bibnamefont {Ghasemkhani}}, \bibinfo
  {author} {\bibfnamefont {D.~V.}\ \bibnamefont {Seletskiy}}, \bibinfo {author}
  {\bibfnamefont {J.~G.}\ \bibnamefont {Cederberg}}, \ and\ \bibinfo {author}
  {\bibfnamefont {M.}~\bibnamefont {Sheik-Bahae}},\ }\bibfield  {title}
  {\enquote {\bibinfo {title} {{Exploring ultrafast negative Kerr effect for
  mode-locking vertical external-cavity surface-emitting lasers}},}\
  }\href@noop {} {\bibfield  {journal} {\bibinfo  {journal} {Optics Express}\
  }\textbf {\bibinfo {volume} {21}},\ \bibinfo {pages} {28801--28808} (\bibinfo
  {year} {2013})}\BibitemShut {NoStop}%
\bibitem [{\citenamefont {Gaafar}\ \emph {et~al.}(2014)\citenamefont {Gaafar},
  \citenamefont {Richter}, \citenamefont {Keskin}, \citenamefont
  {M{\"{o}}ller}, \citenamefont {Wichmann}, \citenamefont {Stolz},
  \citenamefont {Rahimi-Iman},\ and\ \citenamefont {Koch}}]{Gaafar2014a}%
  \BibitemOpen
  \bibfield  {author} {\bibinfo {author} {\bibfnamefont {M.}~\bibnamefont
  {Gaafar}}, \bibinfo {author} {\bibfnamefont {P.}~\bibnamefont {Richter}},
  \bibinfo {author} {\bibfnamefont {H.}~\bibnamefont {Keskin}}, \bibinfo
  {author} {\bibfnamefont {C.}~\bibnamefont {M{\"{o}}ller}}, \bibinfo {author}
  {\bibfnamefont {M.}~\bibnamefont {Wichmann}}, \bibinfo {author}
  {\bibfnamefont {W.}~\bibnamefont {Stolz}}, \bibinfo {author} {\bibfnamefont
  {A.}~\bibnamefont {Rahimi-Iman}}, \ and\ \bibinfo {author} {\bibfnamefont
  {M.}~\bibnamefont {Koch}},\ }\bibfield  {title} {\enquote {\bibinfo {title}
  {{Self-mode-locking semiconductor disk laser}},}\ }\href@noop {} {\bibfield
  {journal} {\bibinfo  {journal} {Optics Express}\ }\textbf {\bibinfo {volume}
  {22}},\ \bibinfo {pages} {28390--28399} (\bibinfo {year} {2014})}\BibitemShut
  {NoStop}%
\bibitem [{\citenamefont {Bek}\ \emph {et~al.}(2017)\citenamefont {Bek},
  \citenamefont {Gro{\ss}mann}, \citenamefont {Kahle}, \citenamefont {Koch},
  \citenamefont {Rahimi-Iman}, \citenamefont {Jetter},\ and\ \citenamefont
  {Michler}}]{Bek2017}%
  \BibitemOpen
  \bibfield  {author} {\bibinfo {author} {\bibfnamefont {R.}~\bibnamefont
  {Bek}}, \bibinfo {author} {\bibfnamefont {M.}~\bibnamefont {Gro{\ss}mann}},
  \bibinfo {author} {\bibfnamefont {H.}~\bibnamefont {Kahle}}, \bibinfo
  {author} {\bibfnamefont {M.}~\bibnamefont {Koch}}, \bibinfo {author}
  {\bibfnamefont {A.}~\bibnamefont {Rahimi-Iman}}, \bibinfo {author}
  {\bibfnamefont {M.}~\bibnamefont {Jetter}}, \ and\ \bibinfo {author}
  {\bibfnamefont {P.}~\bibnamefont {Michler}},\ }\bibfield  {title} {\enquote
  {\bibinfo {title} {{Self-mode-locked AlGaInP-VECSEL}},}\ }\href@noop {}
  {\bibfield  {journal} {\bibinfo  {journal} {Applied Physics Letters}\
  }\textbf {\bibinfo {volume} {111}},\ \bibinfo {pages} {182105} (\bibinfo
  {year} {2017})}\BibitemShut {NoStop}%
\bibitem [{\citenamefont {Wilcox}\ and\ \citenamefont
  {Tropper}(2013)}]{Wilcox2013b}%
  \BibitemOpen
  \bibfield  {author} {\bibinfo {author} {\bibfnamefont {K.~G.}\ \bibnamefont
  {Wilcox}}\ and\ \bibinfo {author} {\bibfnamefont {A.~C.}\ \bibnamefont
  {Tropper}},\ }\bibfield  {title} {\enquote {\bibinfo {title} {{Comment on
  SESAM-free mode-locked semiconductor disk laser}},}\ }\href@noop {}
  {\bibfield  {journal} {\bibinfo  {journal} {Laser and Photonics Reviews}\
  }\textbf {\bibinfo {volume} {423}},\ \bibinfo {pages} {422--423} (\bibinfo
  {year} {2013})}\BibitemShut {NoStop}%
\bibitem [{\citenamefont {Gaafar}\ \emph {et~al.}(2016)\citenamefont {Gaafar},
  \citenamefont {Rahimi-Iman}, \citenamefont {Fedorova}, \citenamefont {Stolz},
  \citenamefont {Rafailov},\ and\ \citenamefont {Koch}}]{Gaafar2016}%
  \BibitemOpen
  \bibfield  {author} {\bibinfo {author} {\bibfnamefont {M.~A.}\ \bibnamefont
  {Gaafar}}, \bibinfo {author} {\bibfnamefont {A.}~\bibnamefont {Rahimi-Iman}},
  \bibinfo {author} {\bibfnamefont {K.~A.}\ \bibnamefont {Fedorova}}, \bibinfo
  {author} {\bibfnamefont {W.}~\bibnamefont {Stolz}}, \bibinfo {author}
  {\bibfnamefont {E.~U.}\ \bibnamefont {Rafailov}}, \ and\ \bibinfo {author}
  {\bibfnamefont {M.}~\bibnamefont {Koch}},\ }\bibfield  {title} {\enquote
  {\bibinfo {title} {{Mode-locked Semiconductor Disk Lasers}},}\ }\href@noop {}
  {\bibfield  {journal} {\bibinfo  {journal} {Advances in Optics and
  Photonics}\ }\textbf {\bibinfo {volume} {8}},\ \bibinfo {pages} {370--400}
  (\bibinfo {year} {2016})}\BibitemShut {NoStop}%
\bibitem [{\citenamefont {Escoto}\ and\ \citenamefont
  {Steinmeyer}(2020)}]{Escoto2020}%
  \BibitemOpen
  \bibfield  {author} {\bibinfo {author} {\bibfnamefont {E.}~\bibnamefont
  {Escoto}}\ and\ \bibinfo {author} {\bibfnamefont {G.}~\bibnamefont
  {Steinmeyer}},\ }\bibfield  {title} {\enquote {\bibinfo {title} {{Pseudo
  Mode-Locking}},}\ }\href@noop {} {\bibfield  {journal} {\bibinfo  {journal}
  {Proc. SPIE 11263, Vertical External Cavity Surface Emitting Lasers (VECSELs)
  X}\ }\textbf {\bibinfo {volume} {11263}},\ \bibinfo {pages} {1--3} (\bibinfo
  {year} {2020})}\BibitemShut {NoStop}%
\bibitem [{\citenamefont {Quarterman}, \citenamefont {Tyrk},\ and\
  \citenamefont {Wilcox}(2015)}]{Quarterman2015}%
  \BibitemOpen
  \bibfield  {author} {\bibinfo {author} {\bibfnamefont {A.~H.}\ \bibnamefont
  {Quarterman}}, \bibinfo {author} {\bibfnamefont {M.~A.}\ \bibnamefont
  {Tyrk}}, \ and\ \bibinfo {author} {\bibfnamefont {K.~G.}\ \bibnamefont
  {Wilcox}},\ }\bibfield  {title} {\enquote {\bibinfo {title} {{Z-scan
  measurements of the nonlinear refractive index of a pumped semiconductor disk
  laser gain medium}},}\ }\href@noop {} {\bibfield  {journal} {\bibinfo
  {journal} {Applied Physics Letters}\ }\textbf {\bibinfo {volume} {106}},\
  \bibinfo {pages} {011105} (\bibinfo {year} {2015})}\BibitemShut {NoStop}%
\bibitem [{\citenamefont {Shaw}\ \emph {et~al.}(2016)\citenamefont {Shaw},
  \citenamefont {Quarterman}, \citenamefont {Turnbull}, \citenamefont {{Chen
  Sverre}}, \citenamefont {Head}, \citenamefont {Tropper},\ and\ \citenamefont
  {Wilcox}}]{Shaw2016}%
  \BibitemOpen
  \bibfield  {author} {\bibinfo {author} {\bibfnamefont {E.~A.}\ \bibnamefont
  {Shaw}}, \bibinfo {author} {\bibfnamefont {A.~H.}\ \bibnamefont
  {Quarterman}}, \bibinfo {author} {\bibfnamefont {A.~P.}\ \bibnamefont
  {Turnbull}}, \bibinfo {author} {\bibfnamefont {T.}~\bibnamefont {{Chen
  Sverre}}}, \bibinfo {author} {\bibfnamefont {C.~R.}\ \bibnamefont {Head}},
  \bibinfo {author} {\bibfnamefont {A.~C.}\ \bibnamefont {Tropper}}, \ and\
  \bibinfo {author} {\bibfnamefont {K.~G.}\ \bibnamefont {Wilcox}},\ }\bibfield
   {title} {\enquote {\bibinfo {title} {{Nonlinear Lensing in an Unpumped
  Antiresonant Semiconductor Disk Laser Gain Structure}},}\ }\href@noop {}
  {\bibfield  {journal} {\bibinfo  {journal} {IEEE Photonics Technology
  Letters}\ }\textbf {\bibinfo {volume} {28}},\ \bibinfo {pages} {1395--1398}
  (\bibinfo {year} {2016})}\BibitemShut {NoStop}%
\bibitem [{\citenamefont {Quarterman}\ \emph {et~al.}(2016)\citenamefont
  {Quarterman}, \citenamefont {Mirkhanov}, \citenamefont {Smyth},\ and\
  \citenamefont {Wilcox}}]{Quarterman2016}%
  \BibitemOpen
  \bibfield  {author} {\bibinfo {author} {\bibfnamefont {A.~H.}\ \bibnamefont
  {Quarterman}}, \bibinfo {author} {\bibfnamefont {S.}~\bibnamefont
  {Mirkhanov}}, \bibinfo {author} {\bibfnamefont {C.~J.}\ \bibnamefont
  {Smyth}}, \ and\ \bibinfo {author} {\bibfnamefont {K.~G.}\ \bibnamefont
  {Wilcox}},\ }\bibfield  {title} {\enquote {\bibinfo {title} {{Measurements of
  nonlinear lensing in a semiconductor disk laser gain sample under optical
  pumping and using a resonant femtosecond probe laser}},}\ }\href@noop {}
  {\bibfield  {journal} {\bibinfo  {journal} {Applied Physics Letters}\
  }\textbf {\bibinfo {volume} {109}},\ \bibinfo {pages} {121113} (\bibinfo
  {year} {2016})}\BibitemShut {NoStop}%
\bibitem [{\citenamefont {Kriso}\ \emph {et~al.}(2019)\citenamefont {Kriso},
  \citenamefont {Kress}, \citenamefont {Munshi}, \citenamefont {Gro{\ss}mann},
  \citenamefont {Bek}, \citenamefont {Jetter}, \citenamefont {Michler},
  \citenamefont {Stolz}, \citenamefont {Koch},\ and\ \citenamefont
  {Rahimi-Iman}}]{Kriso2019}%
  \BibitemOpen
  \bibfield  {author} {\bibinfo {author} {\bibfnamefont {C.}~\bibnamefont
  {Kriso}}, \bibinfo {author} {\bibfnamefont {S.}~\bibnamefont {Kress}},
  \bibinfo {author} {\bibfnamefont {T.}~\bibnamefont {Munshi}}, \bibinfo
  {author} {\bibfnamefont {M.}~\bibnamefont {Gro{\ss}mann}}, \bibinfo {author}
  {\bibfnamefont {R.}~\bibnamefont {Bek}}, \bibinfo {author} {\bibfnamefont
  {M.}~\bibnamefont {Jetter}}, \bibinfo {author} {\bibfnamefont
  {P.}~\bibnamefont {Michler}}, \bibinfo {author} {\bibfnamefont
  {W.}~\bibnamefont {Stolz}}, \bibinfo {author} {\bibfnamefont
  {M.}~\bibnamefont {Koch}}, \ and\ \bibinfo {author} {\bibfnamefont
  {A.}~\bibnamefont {Rahimi-Iman}},\ }\bibfield  {title} {\enquote {\bibinfo
  {title} {{Microcavity-enhanced Kerr nonlinearity in a
  vertical-external-cavity surface-emitting laser}},}\ }\href@noop {}
  {\bibfield  {journal} {\bibinfo  {journal} {Optics Express}\ }\textbf
  {\bibinfo {volume} {27}},\ \bibinfo {pages} {11914--11929} (\bibinfo {year}
  {2019})}\BibitemShut {NoStop}%
\bibitem [{\citenamefont {Kriso}\ \emph {et~al.}(2020)\citenamefont {Kriso},
  \citenamefont {Kress}, \citenamefont {Munshi}, \citenamefont {Grossmann},
  \citenamefont {Bek}, \citenamefont {Jetter}, \citenamefont {Michler},
  \citenamefont {Stolz}, \citenamefont {Koch},\ and\ \citenamefont
  {Rahimi-Iman}}]{Kriso2020}%
  \BibitemOpen
  \bibfield  {author} {\bibinfo {author} {\bibfnamefont {C.}~\bibnamefont
  {Kriso}}, \bibinfo {author} {\bibfnamefont {S.}~\bibnamefont {Kress}},
  \bibinfo {author} {\bibfnamefont {T.}~\bibnamefont {Munshi}}, \bibinfo
  {author} {\bibfnamefont {M.}~\bibnamefont {Grossmann}}, \bibinfo {author}
  {\bibfnamefont {R.}~\bibnamefont {Bek}}, \bibinfo {author} {\bibfnamefont
  {M.}~\bibnamefont {Jetter}}, \bibinfo {author} {\bibfnamefont
  {P.}~\bibnamefont {Michler}}, \bibinfo {author} {\bibfnamefont
  {W.}~\bibnamefont {Stolz}}, \bibinfo {author} {\bibfnamefont
  {M.}~\bibnamefont {Koch}}, \ and\ \bibinfo {author} {\bibfnamefont
  {A.}~\bibnamefont {Rahimi-Iman}},\ }\bibfield  {title} {\enquote {\bibinfo
  {title} {{Wavelength and Pump-Power Dependent Nonlinear Refraction and
  Absorption in a Semiconductor Disk Laser}},}\ }\href@noop {} {\bibfield
  {journal} {\bibinfo  {journal} {IEEE Photonics Technology Letters}\ }\textbf
  {\bibinfo {volume} {32}},\ \bibinfo {pages} {85--88} (\bibinfo {year}
  {2020})}\BibitemShut {NoStop}%
\bibitem [{\citenamefont {Paschotta}\ \emph {et~al.}(2002)\citenamefont
  {Paschotta}, \citenamefont {H{\"{a}}ring}, \citenamefont {Garnache},
  \citenamefont {Hoogland}, \citenamefont {Tropper},\ and\ \citenamefont
  {Keller}}]{Paschotta2002}%
  \BibitemOpen
  \bibfield  {author} {\bibinfo {author} {\bibfnamefont {R.}~\bibnamefont
  {Paschotta}}, \bibinfo {author} {\bibfnamefont {R.}~\bibnamefont
  {H{\"{a}}ring}}, \bibinfo {author} {\bibfnamefont {A.}~\bibnamefont
  {Garnache}}, \bibinfo {author} {\bibfnamefont {S.}~\bibnamefont {Hoogland}},
  \bibinfo {author} {\bibfnamefont {A.~C.}\ \bibnamefont {Tropper}}, \ and\
  \bibinfo {author} {\bibfnamefont {U.}~\bibnamefont {Keller}},\ }\bibfield
  {title} {\enquote {\bibinfo {title} {{Soliton-like pulse-shaping mechanism in
  passively mode-locked surface-emitting semiconductor lasers}},}\ }\href@noop
  {} {\bibfield  {journal} {\bibinfo  {journal} {Applied Physics B: Lasers and
  Optics}\ }\textbf {\bibinfo {volume} {75}},\ \bibinfo {pages} {445--451}
  (\bibinfo {year} {2002})}\BibitemShut {NoStop}%
\bibitem [{\citenamefont {Sieber}\ \emph {et~al.}(2013)\citenamefont {Sieber},
  \citenamefont {Hoffmann}, \citenamefont {Wittwer}, \citenamefont {Mangold},
  \citenamefont {Golling}, \citenamefont {Tilma}, \citenamefont
  {S{\"{u}}dmeyer},\ and\ \citenamefont {Keller}}]{Sieber2013}%
  \BibitemOpen
  \bibfield  {author} {\bibinfo {author} {\bibfnamefont {O.~D.}\ \bibnamefont
  {Sieber}}, \bibinfo {author} {\bibfnamefont {M.}~\bibnamefont {Hoffmann}},
  \bibinfo {author} {\bibfnamefont {V.~J.}\ \bibnamefont {Wittwer}}, \bibinfo
  {author} {\bibfnamefont {M.}~\bibnamefont {Mangold}}, \bibinfo {author}
  {\bibfnamefont {M.}~\bibnamefont {Golling}}, \bibinfo {author} {\bibfnamefont
  {B.~W.}\ \bibnamefont {Tilma}}, \bibinfo {author} {\bibfnamefont
  {T.}~\bibnamefont {S{\"{u}}dmeyer}}, \ and\ \bibinfo {author} {\bibfnamefont
  {U.}~\bibnamefont {Keller}},\ }\bibfield  {title} {\enquote {\bibinfo {title}
  {{Experimentally verified pulse formation model for high-power femtosecond
  VECSELs}},}\ }\href@noop {} {\bibfield  {journal} {\bibinfo  {journal}
  {Applied Physics B: Lasers and Optics}\ }\textbf {\bibinfo {volume} {113}},\
  \bibinfo {pages} {133--145} (\bibinfo {year} {2013})}\BibitemShut {NoStop}%
\bibitem [{\citenamefont {Ferdinandus}\ \emph {et~al.}(2013)\citenamefont
  {Ferdinandus}, \citenamefont {Hu}, \citenamefont {Reichert}, \citenamefont
  {Wang}, \citenamefont {Hagan},\ and\ \citenamefont
  {Stryland}}]{Ferdinandus2013}%
  \BibitemOpen
  \bibfield  {author} {\bibinfo {author} {\bibfnamefont {M.~R.}\ \bibnamefont
  {Ferdinandus}}, \bibinfo {author} {\bibfnamefont {H.}~\bibnamefont {Hu}},
  \bibinfo {author} {\bibfnamefont {M.}~\bibnamefont {Reichert}}, \bibinfo
  {author} {\bibfnamefont {Z.}~\bibnamefont {Wang}}, \bibinfo {author}
  {\bibfnamefont {D.~J.}\ \bibnamefont {Hagan}}, \ and\ \bibinfo {author}
  {\bibfnamefont {E.~W.}\ \bibnamefont {Stryland}},\ }\bibfield  {title}
  {\enquote {\bibinfo {title} {{Beam deflection measurement of time and
  polarization resolved nonlinear refraction}},}\ }\href@noop {} {\bibfield
  {journal} {\bibinfo  {journal} {Optics Letters}\ }\textbf {\bibinfo {volume}
  {38}},\ \bibinfo {pages} {3518--3521} (\bibinfo {year} {2013})}\BibitemShut
  {NoStop}%
\bibitem [{\citenamefont {Sheik-Bahae}\ \emph {et~al.}(1990)\citenamefont
  {Sheik-Bahae}, \citenamefont {Said}, \citenamefont {Wei}, \citenamefont
  {Hagan},\ and\ \citenamefont {{Van Stryland}}}]{Sheik-Bahae1990}%
  \BibitemOpen
  \bibfield  {author} {\bibinfo {author} {\bibfnamefont {M.}~\bibnamefont
  {Sheik-Bahae}}, \bibinfo {author} {\bibfnamefont {A.~A.}\ \bibnamefont
  {Said}}, \bibinfo {author} {\bibfnamefont {T.~H.}\ \bibnamefont {Wei}},
  \bibinfo {author} {\bibfnamefont {D.~J.}\ \bibnamefont {Hagan}}, \ and\
  \bibinfo {author} {\bibfnamefont {E.~W.}\ \bibnamefont {{Van Stryland}}},\
  }\bibfield  {title} {\enquote {\bibinfo {title} {{Sensitive Measurement of
  Optical Nonlinearities Using a Single Beam}},}\ }\href@noop {} {\bibfield
  {journal} {\bibinfo  {journal} {IEEE Journal of Quantum Electronics}\
  }\textbf {\bibinfo {volume} {26}},\ \bibinfo {pages} {760--769} (\bibinfo
  {year} {1990})}\BibitemShut {NoStop}%
\bibitem [{\citenamefont {Yang}\ \emph {et~al.}(2015)\citenamefont {Yang},
  \citenamefont {Albrecht}, \citenamefont {Cederberg},\ and\ \citenamefont
  {Sheik-Bahae}}]{Yang2015}%
  \BibitemOpen
  \bibfield  {author} {\bibinfo {author} {\bibfnamefont {Z.}~\bibnamefont
  {Yang}}, \bibinfo {author} {\bibfnamefont {A.~R.}\ \bibnamefont {Albrecht}},
  \bibinfo {author} {\bibfnamefont {J.~G.}\ \bibnamefont {Cederberg}}, \ and\
  \bibinfo {author} {\bibfnamefont {M.}~\bibnamefont {Sheik-Bahae}},\
  }\bibfield  {title} {\enquote {\bibinfo {title} {{Optically pumped DBR-free
  semiconductor disk lasers}},}\ }\href@noop {} {\bibfield  {journal} {\bibinfo
   {journal} {Optics Express}\ }\textbf {\bibinfo {volume} {23}},\ \bibinfo
  {pages} {2419--2421} (\bibinfo {year} {2015})}\BibitemShut {NoStop}%
\bibitem [{\citenamefont {Kahle}\ \emph {et~al.}(2016)\citenamefont {Kahle},
  \citenamefont {Mateo}, \citenamefont {Brauch}, \citenamefont {Tatar-Mathes},
  \citenamefont {Bek}, \citenamefont {Jetter}, \citenamefont {Graf},\ and\
  \citenamefont {Michler}}]{Kahle2016}%
  \BibitemOpen
  \bibfield  {author} {\bibinfo {author} {\bibfnamefont {H.}~\bibnamefont
  {Kahle}}, \bibinfo {author} {\bibfnamefont {C.~M.~N.}\ \bibnamefont {Mateo}},
  \bibinfo {author} {\bibfnamefont {U.}~\bibnamefont {Brauch}}, \bibinfo
  {author} {\bibfnamefont {P.}~\bibnamefont {Tatar-Mathes}}, \bibinfo {author}
  {\bibfnamefont {R.}~\bibnamefont {Bek}}, \bibinfo {author} {\bibfnamefont
  {M.}~\bibnamefont {Jetter}}, \bibinfo {author} {\bibfnamefont
  {T.}~\bibnamefont {Graf}}, \ and\ \bibinfo {author} {\bibfnamefont
  {P.}~\bibnamefont {Michler}},\ }\bibfield  {title} {\enquote {\bibinfo
  {title} {{Semiconductor membrane external-cavity surface-emitting laser
  (MECSEL)}},}\ }\href {\doibase 10.1364/optica.3.001506} {\bibfield  {journal}
  {\bibinfo  {journal} {Optica}\ }\textbf {\bibinfo {volume} {3}},\ \bibinfo
  {pages} {1506--1512} (\bibinfo {year} {2016})}\BibitemShut {NoStop}%
\bibitem [{\citenamefont {Yang}\ \emph {et~al.}(2018)\citenamefont {Yang},
  \citenamefont {Follman}, \citenamefont {Albrecht}, \citenamefont {Heu},
  \citenamefont {Giannini}, \citenamefont {Cole},\ and\ \citenamefont
  {Sheik-Bahae}}]{Yang2018}%
  \BibitemOpen
  \bibfield  {author} {\bibinfo {author} {\bibfnamefont {Z.}~\bibnamefont
  {Yang}}, \bibinfo {author} {\bibfnamefont {D.}~\bibnamefont {Follman}},
  \bibinfo {author} {\bibfnamefont {A.~R.}\ \bibnamefont {Albrecht}}, \bibinfo
  {author} {\bibfnamefont {P.}~\bibnamefont {Heu}}, \bibinfo {author}
  {\bibfnamefont {N.}~\bibnamefont {Giannini}}, \bibinfo {author}
  {\bibfnamefont {G.~D.}\ \bibnamefont {Cole}}, \ and\ \bibinfo {author}
  {\bibfnamefont {M.}~\bibnamefont {Sheik-Bahae}},\ }\bibfield  {title}
  {\enquote {\bibinfo {title} {{16 W DBR-free membrane semiconductor disk laser
  with dual-SiC heatspreader}},}\ }\href@noop {} {\bibfield  {journal}
  {\bibinfo  {journal} {Electronics Letters}\ }\textbf {\bibinfo {volume}
  {54}},\ \bibinfo {pages} {430--432} (\bibinfo {year} {2018})}\BibitemShut
  {NoStop}%
\bibitem [{\citenamefont {Hall}\ \emph {et~al.}(1990)\citenamefont {Hall},
  \citenamefont {Mark}, \citenamefont {Ippen},\ and\ \citenamefont
  {Eisenstein}}]{Hall1990}%
  \BibitemOpen
  \bibfield  {author} {\bibinfo {author} {\bibfnamefont {K.}~\bibnamefont
  {Hall}}, \bibinfo {author} {\bibfnamefont {J.}~\bibnamefont {Mark}}, \bibinfo
  {author} {\bibfnamefont {E.}~\bibnamefont {Ippen}}, \ and\ \bibinfo {author}
  {\bibfnamefont {G.}~\bibnamefont {Eisenstein}},\ }\bibfield  {title}
  {\enquote {\bibinfo {title} {{Femtosecond gain dynamics in InGaAsP optical
  amplifiers}},}\ }\href@noop {} {\bibfield  {journal} {\bibinfo  {journal}
  {Applied Physics Letters}\ }\textbf {\bibinfo {volume} {56}},\ \bibinfo
  {pages} {1740--1742} (\bibinfo {year} {1990})}\BibitemShut {NoStop}%
\bibitem [{\citenamefont {Mork}\ and\ \citenamefont
  {Mecozzi}(1994)}]{Mork1994}%
  \BibitemOpen
  \bibfield  {author} {\bibinfo {author} {\bibfnamefont {J.}~\bibnamefont
  {Mork}}\ and\ \bibinfo {author} {\bibfnamefont {A.}~\bibnamefont {Mecozzi}},\
  }\bibfield  {title} {\enquote {\bibinfo {title} {{Response function for gain
  and refractive index dynamics in active semiconductor waveguides}},}\
  }\href@noop {} {\bibfield  {journal} {\bibinfo  {journal} {Applied Physics
  Letters}\ }\textbf {\bibinfo {volume} {65}},\ \bibinfo {pages} {1736--1738}
  (\bibinfo {year} {1994})}\BibitemShut {NoStop}%
\bibitem [{\citenamefont {Alfieri}\ \emph {et~al.}(2017)\citenamefont
  {Alfieri}, \citenamefont {Waldburger}, \citenamefont {Link}, \citenamefont
  {Gini}, \citenamefont {Golling}, \citenamefont {Eisenstein},\ and\
  \citenamefont {Keller}}]{Alfieri2017}%
  \BibitemOpen
  \bibfield  {author} {\bibinfo {author} {\bibfnamefont {C.~G.~E.}\
  \bibnamefont {Alfieri}}, \bibinfo {author} {\bibfnamefont {D.}~\bibnamefont
  {Waldburger}}, \bibinfo {author} {\bibfnamefont {S.~M.}\ \bibnamefont
  {Link}}, \bibinfo {author} {\bibfnamefont {E.}~\bibnamefont {Gini}}, \bibinfo
  {author} {\bibfnamefont {M.}~\bibnamefont {Golling}}, \bibinfo {author}
  {\bibfnamefont {G.}~\bibnamefont {Eisenstein}}, \ and\ \bibinfo {author}
  {\bibfnamefont {U.}~\bibnamefont {Keller}},\ }\bibfield  {title} {\enquote
  {\bibinfo {title} {{Optical efficiency and gain dynamics of modelocked
  semiconductor disk lasers}},}\ }\href@noop {} {\bibfield  {journal} {\bibinfo
   {journal} {Optics Express}\ }\textbf {\bibinfo {volume} {25}},\ \bibinfo
  {pages} {2719--2721} (\bibinfo {year} {2017})}\BibitemShut {NoStop}%
\bibitem [{\citenamefont {Baker}\ \emph {et~al.}(2015)\citenamefont {Baker},
  \citenamefont {Scheller}, \citenamefont {Koch}, \citenamefont {Perez},
  \citenamefont {Stolz}, \citenamefont {{Jason Jones}},\ and\ \citenamefont
  {Moloney}}]{Baker2015}%
  \BibitemOpen
  \bibfield  {author} {\bibinfo {author} {\bibfnamefont {C.}~\bibnamefont
  {Baker}}, \bibinfo {author} {\bibfnamefont {M.}~\bibnamefont {Scheller}},
  \bibinfo {author} {\bibfnamefont {S.~W.}\ \bibnamefont {Koch}}, \bibinfo
  {author} {\bibfnamefont {A.~R.}\ \bibnamefont {Perez}}, \bibinfo {author}
  {\bibfnamefont {W.}~\bibnamefont {Stolz}}, \bibinfo {author} {\bibfnamefont
  {R.}~\bibnamefont {{Jason Jones}}}, \ and\ \bibinfo {author} {\bibfnamefont
  {J.~V.}\ \bibnamefont {Moloney}},\ }\bibfield  {title} {\enquote {\bibinfo
  {title} {{In situ probing of mode-locked
  vertical-external-cavity-surface-emitting lasers}},}\ }\href@noop {}
  {\bibfield  {journal} {\bibinfo  {journal} {Optics Letters}\ }\textbf
  {\bibinfo {volume} {40}},\ \bibinfo {pages} {5459--5462} (\bibinfo {year}
  {2015})}\BibitemShut {NoStop}%
\bibitem [{\citenamefont {Mork}, \citenamefont {Mecozzi},\ and\ \citenamefont
  {Hultgren}(1995)}]{Mork1995}%
  \BibitemOpen
  \bibfield  {author} {\bibinfo {author} {\bibfnamefont {J.}~\bibnamefont
  {Mork}}, \bibinfo {author} {\bibfnamefont {A.}~\bibnamefont {Mecozzi}}, \
  and\ \bibinfo {author} {\bibfnamefont {C.}~\bibnamefont {Hultgren}},\
  }\bibfield  {title} {\enquote {\bibinfo {title} {{Spectral effects in short
  pulse pump-probe measurements}},}\ }\href@noop {} {\bibfield  {journal}
  {\bibinfo  {journal} {Applied Physics Letters}\ }\textbf {\bibinfo {volume}
  {68}},\ \bibinfo {pages} {449--451} (\bibinfo {year} {1995})}\BibitemShut
  {NoStop}%
\bibitem [{\citenamefont {Reichert}\ \emph {et~al.}(2014)\citenamefont
  {Reichert}, \citenamefont {Honghua}, \citenamefont {Ferdinandus},
  \citenamefont {Seidel}, \citenamefont {Zhao}, \citenamefont {Ensley},
  \citenamefont {Peceli}, \citenamefont {Reed}, \citenamefont {Fishman},
  \citenamefont {Webster}, \citenamefont {Hagan},\ and\ \citenamefont {{Van
  Stryland}}}]{Reichert2014}%
  \BibitemOpen
  \bibfield  {author} {\bibinfo {author} {\bibfnamefont {M.}~\bibnamefont
  {Reichert}}, \bibinfo {author} {\bibfnamefont {H.}~\bibnamefont {Honghua}},
  \bibinfo {author} {\bibfnamefont {M.~R.}\ \bibnamefont {Ferdinandus}},
  \bibinfo {author} {\bibfnamefont {M.}~\bibnamefont {Seidel}}, \bibinfo
  {author} {\bibfnamefont {P.}~\bibnamefont {Zhao}}, \bibinfo {author}
  {\bibfnamefont {T.~R.}\ \bibnamefont {Ensley}}, \bibinfo {author}
  {\bibfnamefont {D.}~\bibnamefont {Peceli}}, \bibinfo {author} {\bibfnamefont
  {J.~M.}\ \bibnamefont {Reed}}, \bibinfo {author} {\bibfnamefont {D.~A.}\
  \bibnamefont {Fishman}}, \bibinfo {author} {\bibfnamefont {S.}~\bibnamefont
  {Webster}}, \bibinfo {author} {\bibfnamefont {D.~J.}\ \bibnamefont {Hagan}},
  \ and\ \bibinfo {author} {\bibfnamefont {E.~W.}\ \bibnamefont {{Van
  Stryland}}},\ }\bibfield  {title} {\enquote {\bibinfo {title} {{Temporal,
  spectral, and polarization dependence of the nonlinear optical response of
  carbon disulfide}},}\ }\href@noop {} {\bibfield  {journal} {\bibinfo
  {journal} {Optica}\ }\textbf {\bibinfo {volume} {1}},\ \bibinfo {pages}
  {436--445} (\bibinfo {year} {2014})}\BibitemShut {NoStop}%
\bibitem [{\citenamefont {Sheik-Bahae}\ and\ \citenamefont {{Van
  Stryland}}(1994)}]{Sheik-Bahae1994}%
  \BibitemOpen
  \bibfield  {author} {\bibinfo {author} {\bibfnamefont {M.}~\bibnamefont
  {Sheik-Bahae}}\ and\ \bibinfo {author} {\bibfnamefont {E.}~\bibnamefont {{Van
  Stryland}}},\ }\bibfield  {title} {\enquote {\bibinfo {title} {{Ultrafast
  nonlinearities in semiconductor laser amplifiers}},}\ }\href@noop {}
  {\bibfield  {journal} {\bibinfo  {journal} {Physical Review B}\ }\textbf
  {\bibinfo {volume} {50}},\ \bibinfo {pages} {171--178} (\bibinfo {year}
  {1994})}\BibitemShut {NoStop}%
\bibitem [{\citenamefont {Rahimi-Iman}\ \emph {et~al.}(2016)\citenamefont
  {Rahimi-Iman}, \citenamefont {Gaafar}, \citenamefont {M{\"{o}}ller},
  \citenamefont {Vaupel}, \citenamefont {Zhang}, \citenamefont {Al-nakdali},
  \citenamefont {Fedorova}, \citenamefont {Stolz}, \citenamefont {Rafailov},\
  and\ \citenamefont {Koch}}]{Rahimi-Iman2016b}%
  \BibitemOpen
  \bibfield  {author} {\bibinfo {author} {\bibfnamefont {A.}~\bibnamefont
  {Rahimi-Iman}}, \bibinfo {author} {\bibfnamefont {M.}~\bibnamefont {Gaafar}},
  \bibinfo {author} {\bibfnamefont {C.}~\bibnamefont {M{\"{o}}ller}}, \bibinfo
  {author} {\bibfnamefont {M.}~\bibnamefont {Vaupel}}, \bibinfo {author}
  {\bibfnamefont {F.}~\bibnamefont {Zhang}}, \bibinfo {author} {\bibfnamefont
  {D.}~\bibnamefont {Al-nakdali}}, \bibinfo {author} {\bibfnamefont {K.~A.}\
  \bibnamefont {Fedorova}}, \bibinfo {author} {\bibfnamefont {W.}~\bibnamefont
  {Stolz}}, \bibinfo {author} {\bibfnamefont {E.~U.}\ \bibnamefont {Rafailov}},
  \ and\ \bibinfo {author} {\bibfnamefont {M.}~\bibnamefont {Koch}},\
  }\bibfield  {title} {\enquote {\bibinfo {title} {{Self-Mode-Locked
  Vertical-External-Cavity Surface-Emitting Laser}},}\ }in\ \href@noop {}
  {\emph {\bibinfo {booktitle} {Proc. SPIE 9734, Vertical External Cavity
  Surface Emitting Lasers (VECSELs) VI, 97340M}}}\ (\bibinfo {year}
  {2016})\BibitemShut {NoStop}%
\end{thebibliography}%
\end{document}